\documentclass[12pt]{article}

\def\be{\begin{equation}}
\def\ee{\end{equation}}
\def\bea{\begin{eqnarray}}
\def\eea{\end{eqnarray}}

\def\3vd{\rangle{\hspace{-0.18em}\longrightarrow}{\hspace{-0.65em}^\mid}}
\def\4vd{\rangle{\hspace{-0.15em}\longrightarrow}}

\def\2d{\dot{2}}
\def\1d{\dot{1}}
\def\3d{\dot{3}}

\def\kk{\kappa}

\def\m{\mu}
\def\n{\nu}

\def\1q{Q+1}
\def\2q{Q+\kk(1,2)}
\def\3q{Q+\kk(1,3)}

\def\trr{\triangleright}
\def\p{\partial}
\def\a{\alpha}
\def\b{\beta}
\def\t{\theta}

\usepackage{graphicx}
\usepackage{bm}
\usepackage{amsmath}
\usepackage{amssymb}

\begin{document}


\pagestyle{plain}

\begin{center}
\vspace{2.5cm}

{\Large {\bf Electrodynamics in Noncommutative Curved Space Time}}

\vspace{.5cm}
{\it Abolfazl Jafari}

\vspace{.5cm}
{\it Department of Science, Shahrekord University}

\vspace{.5cm}
{\it Shahrekord, P. O. Box 115, Iran}

\vspace{.3cm}
\texttt{jafari-ab@sci.sku.ac.ir}

\vskip .5 cm
\end{center}
$\mathbf{Abstract.}$ We study the issue of the electrodynamics theory in noncommutative curved space time (NCCST)
with a new star-product.
In this paper, the motion equation of
electrodynamics
and canonical energy-momentum tensor in noncommutative curved space time
will be found.
The most important point is the assumption of the noncommutative parameter ($\theta$) be $x^{\m}$-independent.

\vspace{2cm}

\date{\today}

\newpage

$$\mathbf{Introduction}$$
Since Newton the concept of space time has gone through various
changes. All stages, however, had in common the notion of a continues linear
space. Today we formulate fundamental laws of physics, field theories, gauge field
theories, and the theory of gravity on differentiable manifolds.
That a change in the
concept of space for very short distances might be necessary was already anticipated
by Riemann.
There are indications today
that at very short distances we might have to go beyond differential manifolds. This
is only one of several arguments that we have to expect some changes in physics
for very short distances.
Other arguments are based on the singularity problem in
quantum field theory and the fact theory of gravity is non re-normalizable when
quantized. Why not try an algebraic concept of space time that could guide us to
changes in our present formulation of laws of physics? This is different from the
discovery of quantum mechanics.
There physics data forced us to introduce the concept
of noncommutativity.

There are many approaches to noncommutative geometry and its use in physics. The operator algebra
and $\mathcal{C}^\star$-algebra one, the deformation quantization one, the quantum group one, and the matrix algebra
(fuzzy geometry) one. These show that the concepts of noncommutative subjects were born many years ago, where
the idea of noncommutative
coordinates is almost as old as quantum mechanics.

Many people are working in the noncommutative field theory to
find the details of it. It is a nonlocal theory and for this reason,
the NCFT has many ambiguous problems such as UV/IR divergence and causality.
The UV/IR problem arises of non-locality directly and the causality is broken
when $\theta^{0i}\neq0$ so as a field theory it is not appealing \cite{wess,jab}.
In
addition, some people thought that the noncommutative field theory violets the Lorentz
invariance, but a group of physicists have shown that the noncommutative
field theory holds the Lorentz invariance by Hopf algebra \cite{jab}. With all these issues,
many of physicists prefer to work in this
field, because they think it is one of the best candidate for the brighter future of
physics and they hope that these problems would be solved in futures.

The formulation of the classical and quantum field theories in NCCST has been
a very active subject over the last few years. Most of these approaches focus on
free or interacting QFTs on the Moyal-Weyl deformed or $\kappa$-deformed Minkowski
space time. But at now, many people are working on the quantum field theory and related them on the noncommutative curved
space time so this work is one of them. Finally, we hope that these works might solve some conceptual problems
of physics that are still left at very short distances \cite{wess,jab,nekrasov,jaf}.

We introduce new star-product briefly. By
deforming the ordinary Moyal $\star$-product, we propose a new Moyal-Weyl
$\trr$-product which takes into consideration the missing terms cited above which generate
gravitational terms to the order $\t^2$. In reference \cite{khallili}, the authors have shown that we can
go to curved space time
with replacing of operators variables.
We can replace the
noncommutative flat space time coordinates variables $[\hat{x}^\m\ ,\ \hat{x}^\n]=\imath\t^{\m\n}$
with noncommutative curved space time coordinates variables
$\hat{X}^\m=\hat{x}^\m+\imath^2\frac{\t^{ab}\t^{\a\b}}{2\sqrt{-\hat{g}}}\p_b \hat{R}^{\m}_{a\a\b}$
where $\hat{R}^\m_{b\a\b}$ stands for Riemann curvature tensor and $\sqrt{-\hat{g}}$ is determinant of metric
as two functions in noncommutative coordinates.
We know
$[\hat{X}^\m\ ,\ \hat{X}^\n]=\imath\t^{\m\n}$ so for two any smooth function A and B we have
$$\hat{A}(\hat{X})\hat{B}(\hat{X})=A(x)\ e^{
\frac{\imath}{2}\t^{\m\n}\overleftarrow{\p}_\m\otimes\overrightarrow{\p}_\n+\imath(\imath^2\frac{\t^{ab}\t^{\a\b}}{2\sqrt{-g}}\p_b R^{\m}_{a\a\b})
\p_\m\ (\Im_A\otimes\Im_B)}\ B(x)$$
where $\Im$ stands for the identity of spaces ( for example $\Im_A$ is identity of "A").
If $\triangle^\m=\imath^2\frac{\t^{ab}\t^{\a\b}}{2\sqrt{-g}}\p_b \hat{R}^{\m}_{a\a\b}$ and
$\trr\equiv e^{\triangle x^\m \p_\m(\Im \otimes\Im)
+\frac{\imath\theta^{\m\n}}{2}\p_\m\otimes \p_\n}$ so we have
$(A\trr B)(\hat{x})=A\ e^{
\frac{\imath}{2}\t^{\m\n}\overleftarrow{\p}_\m\otimes\overrightarrow{\p}_\n+\imath\triangle^\m
\p_\m\ (\Im_A\otimes\Im_B)}\ B$. If the $\triangle$ is $x^\m$-independent we can write
\bea
\int d^{(d-1)} x\ (A\trr B)(\hat{x})&=&\int d^{(d-1)} x\
\int\ \mathbf{dkdq}\tilde{A}(k)\tilde{B}(q)\ e^{\imath(k+q)\cdot\hat{x}
-\frac{\imath}{2}\t^{\m\n}k_\m p_\n+\imath\triangle^\m(k+q)_\m}\nonumber\\&=&
\int d^{(d-1)} x\ e^{\triangle^\m\p_\m}\int\ \mathbf{dkdq}\tilde{A}(k)\tilde{B}(q)\ e^{\imath(k+q)\cdot\hat{x}
-\frac{\imath}{2}\t^{\m\n}k_\m p_\n}\nonumber\\&=&\int d^{(d-1)} x (A(x)\star B(x)+\triangle^\m\p_\m(A(x)B(x))+0(\t^4))
\nonumber\\&=&\int d^{(d-1)} x A(x)\star B(x)
\eea
where $\mathbf{dq}=\frac{1}{(2\pi)^\frac{d-1}{2}} dq$ and
we can remove the total derivative term when $\triangle$ be a constant.
At last, there is a quality below of integral between of old and new star-product for some manifolds with special
metric which $\triangle$ over of them will be a constant.

We introduce a new symbol $S_{\star}(A_1,A_2,..,A_n)$ that it includes all symmetric compounds
of $A_1$-$A_n$.
As a function in integration, because of rotational properties
the one of $A_i$ from $S_{\star}(A_1,A_2,..,A_n)$
can be written in the first (last) of all compounds of $S_\star$ or
$S_{\star}(A_i;A_{i-1},..,A_n,..)$ where $A_i$ is a selected function.
Also, we can show that $\star$-product in integral formalism has following property
$$\int d^d x\ S_{\star}(A_1,A_2,..,A_n)=\int d^d x\ S_{\star}(A_i;A_{i-1},..,A_n,..)=
\int d^d x\ A_i\star \bar{S}^{i}_{\star}(A_{i-1},..,A_n,..)$$ where $\bar{S}^{i}_{\star}$ means $S_\star$ without $A_i$
and includes all symmetric compounds of remanned $A_1,..,A_{i-1},A_{i+1},..,A_n$. The $S_{\star}(A_1,..,A_n)$
has $n!$-terms out of integration but in the integration formalism it reduces to $n$-terms.
Suppose, that $A$-function has appeared twice in $S_\star(B,C,..,A,D,..,A,G,..,H)$, so for
$I=\int d^d x\ S_{\star}(B,C,..,A,D,..,A,G,..,H)$ from
Leibnitz rule we have
\bea
\frac{\delta}{\delta A(z)}I&=&
\int d^d x\ S_{\star}(B,C,..,\frac{\delta}{\delta A(z)}A,D,..,A,G,..,H)\nonumber\\&+&
S_{\star}(B,C,..,A,D,..,\frac{\delta}{\delta A(z)}A,G,..,H)\nonumber\\&=&
\int d^d x\ \frac{\delta}{\delta A(z)}A\star\bar{S}^1_{\star}(D,..,A,G,..,H,B,C,..)
\nonumber\\&+&\int d^d x\ \frac{\delta}{\delta A(z)}A\star\bar{S}^2_{\star}(G,..,H,B,C,..,A,D,..)\nonumber\\&=&
\int d^d x\ \frac{\delta}{\delta A(z)}S^1_{\star}(A;D,..,\acute{A},G,..,H,B,C,..)
\nonumber\\&+&\int d^d x\ \frac{\delta}{\delta A(z)}S^2_{\star}(A;G,..,H,B,C,..,\acute{A},D,..)\nonumber\\&=&
2\bar{S}^{1\ or\ 2}_{\star}
\eea

$$\mathbf{Physics\ \ Notes\ \ and\ the\ \ Motion\ \ Equation\ \ of\ \ Fields}$$

We start by showing how to construct an action electrodynamics in NCCST.
If one direct follows the general rule of transforming usual
theories in noncommutative ones by replacing product of fields by star product \cite{jab,nekrasov}
and we believe that this change should be done on Lagrangian density. In
fact,
$\ \mathbf{S}_{Cm}(\mathcal{L}_{Cm})\rightarrow \mathbf{S}_{Nc}(\mathcal{L}^{Sym}_{Nc})\ $
or
$$\ \mathbf{S}=\int \textit{d}^d x\ \sqrt{-g}\trr {\mathcal{L}_\trr}^{Sym}_{Nc}\ \ $$
where $\trr$ is a new star product.
The classical Lagrangian density is
\bea
\mathcal{L}=\frac{-1}{4}(F_{\m\n}g^{\m\a}g^{\n\b}F_{\a\b})
\eea
where $g^{\m\n}$,$-g$ and $F^{\m\n}$ are the metric tensor, the determinant of metric and the strength fields
$F^{\m\n}=\p^\m A^\n-\p^\n A^\m-\imath\ e [\ A^{\m} , A^\n\ ]_\star$
respectively.
The building of action in NCCST is more complicated
because
symmetric ordering
must also considered.
In particular interested in the deformation of the canonical action goes to
\bea
\mathbf{S}&=&
\int \textit{d}^d x\ \ \frac{-1}{4}\sqrt{-g}\trr
S_\trr(F_{\a\b},g^{\a\m},g^{\b\n},F_{\m\n})
\eea
As for multiplication of the functions we can write $\int\ (f\trr g)(x)=\int\ (f\star g)(\bar{x})$ where $\bar{x}^\m=x^\m+
\triangle^\m$ and if
$\bar{f}=f(\bar{x})$ so then $(f\star g)(\bar{x})=\bar{f}\star\bar {g}$.
The earlier star product ($\star$-product) will be an associative
by the Drinfel'd map. By using these tools, and hermitical structure we have
%
\bea \label{6}&&
S_\trr(F_{\a\b},g^{\a\m},g^{\b\n},F_{\m\n})=
S_\star(\bar{F}_{\a\b},\bar{g}^{\a\m},\bar{g}^{\b\n},\bar{F}_{\m\n})=
\bar{F}\star \bar{g}\star \bar{g}\star \bar{F}
\eea
but we work with $\mathcal{L}_{\star}=S_{\star}(F,g,g,F)$.
Now, we can deform the classical action to the global expression
\bea
\mathbf{S}=
\int \textit{d}^d x\ \sqrt{-\bar{g}}\star \mathcal{L}_\star=
\int \textit{d}^d x\ \sqrt{-g}\star \mathcal{L}_\star+total\ \ derivative
\eea
but when the $\triangle$ be $x$-independent the last term is zero and
%
%
also one of $\star$'s can be
removed $\int f\star g\star h=\int h\star g\star f$.  While the space time without tersion we can use from $F^{\m\n}$.
For research of the motion equation of fields we start from the principle of the least action $\frac{\delta\mathbf{S}}
{\delta A^\kappa(z)}=0$, namely,
%
\bea
\frac{\delta\mathbf{S}}{\delta A^\kappa(z)}&=&\frac{\delta}{\delta A^\kappa(z)}
\int\ \textit{d}^d x\ \
\frac{-\sqrt{-g}}{4}\star S_\star(F,g,g,F)\nonumber\\&=&
\frac{\delta}{\delta A^\kappa(z)}
\int\ \textit{d}^d x\ \
\frac{-1}{4}S_\star(\sqrt{-g};F,g,g,F)\nonumber\\&=&
\frac{\delta}{\delta A^\kappa(z)}
\int\ \textit{d}^d x\ \
\frac{-1}{2}S_\star(F;\sqrt{-g},F,g,g)
\eea
so we have
\bea &&
\p_\m S_{\star}(\sqrt{-g},F_{\a\b},g^{\kappa\a},g^{\m\b})-\imath e [A_\m ,\
S_{\star}(\sqrt{-g},F_{\a\b},g^{\kappa\a},g^{\m\b})]_\star=0
\eea
for Lagranian is diffined in Eq.$\ref{6}$ we have
\bea &&
\p_\m \mathcal{F}^{\kappa\m}-\imath e [A_\m ,\ \mathcal{F}^{\kappa\m} ]=0
\eea
where
$\mathcal{F}^{\kappa\m}=\{ g^{\kappa\b},\ g^{\m\a}\} \star F_{\a\b}\star \sqrt{-g}
+\sqrt{-g}\star F_{\a\b}\star \{ g^{\kappa\b},\ g^{\m\a}\}$
It is the motion equation of electrodynamics fields in NCCST while the $\triangle$ be a constant.


$$\mathbf{the}\ \mathbf{Formal}\ \mathbf{Canonical}\ \mathbf{Energy-momentum}\ \mathbf{Tensor}$$

It is useful to study the derivative the different pieces of the action with respect to $g^{\m\n}$.
In view of the physical interpretations to be given later we introduce the new tensor $\mathbf{T}^{\m\n}$ via
$$\frac{\delta \mathbf{S}}{\delta g_{\m\n}}=-\frac{1}{2}S_\trr(\sqrt{-g}\ ,\ \mathbf{T}^{\m\n})$$
as the field symmetric energy-momentum tensor. Consider the action of field for $\t^{0i}\neq0$.
In this case, we can remove all $\star$'s related to metric tensor, because we do not search the momentum of
metric \cite{Neto2}.
However, for any case of noncommutativity we can start from
\bea
\mathbf{S}^{\t^{0i}=0}=
\int d^d x\sqrt{-g}\trr\mathcal{L}^{\t^{0i}=0}_\trr,\ \
\mathbf{S}^{\t^{0i}\neq0}=\int d^d x\sqrt{-g}\mathcal{L}^{\t^{0i}\neq0}_\trr
\eea
where $\mathcal{L}^{\t^{0i}=0}_\star=S_{\star}(F,g,g,F)$ and
$\mathcal{L}^{\t^{0i}\neq0}_\star=
\frac{-1}{4}g^{\m\a} g^{\n\b}(F_{\m\n}\star F_{\a\b})$.
Variation with respect to $g_{\m\n}$ for two cases give
\bea &&
\delta \mathbf{S}^{\t^{0i}=0}=\int d^d x (\delta \sqrt{-g}\star \mathcal{L}^{\t^{0i}=0}_\star+
\sqrt{-g}\star \delta\mathcal{L}^{\t^{0i}=0}_\star)\nonumber\\&&
\delta \mathbf{S}^{\t^{0i}\neq0}=\int d^d x (\delta \sqrt{-g}\ \mathcal{L}^{\t^{0i}\neq0}_\star+
\sqrt{-g}\ \delta\mathcal{L}^{\t^{0i}\neq0}_\star)
\eea
We first perform the variation of $\sqrt{-g}$ with respect to $\delta g_{\m\n}$ \cite{kleinert}. For this we write
$\delta \sqrt{-g}=-\frac{1}{2\sqrt{-g}}\delta g$ and observe that by
varying $g_{\m\n}(\bar{x})$, the variation of determinant
$g$ involves the co-factors, which in fact are equal to g times the inverse, $g^{\m\n}$ is
$\delta g=-g\ g_{\m\n}\delta g^{\m\n}$
%
%
%
so we have
\bea
\delta\ \sqrt{-g}=\frac{1}{2}S_{\star}(\sqrt{-g},\ g^{\m\n},\ \delta g_{\m\n})=\frac{-1}{2}S_{\star}
(\sqrt{-g},\ g_{\m\n},\ \delta g^{\m\n})
\eea
Therefore, we can write the variation of action
%
\bea
\delta \mathbf{S}^{\t^{0i}\neq0}=\int d^dx ((\frac{-1}{2}S_{\star}(\sqrt{-g},g_{\m\n},\delta g^{\m\n}))\
\mathcal{L}^{\t^{0i}\neq0}_\star+\sqrt{-g}\ \delta\mathcal{L}^{\t^{0i}\neq0}_\star)
\eea
and
\bea
\delta \mathbf{S}^{\t^{0i}=0}=\int d^dx ((\frac{-1}{2}S_{\star}(\sqrt{-g},g_{\m\n},\delta g^{\m\n}))
\star\
\mathcal{L}^{\t^{0i}=0}_\star+\sqrt{-g}\star \delta\mathcal{L}^{\t^{0i}=0}_\star)
\eea
Consider now the variation of Lagrangian density, first case $\t^{0i}\neq0\Rightarrow$
$\delta\mathcal{L}^{\t^{0i}\neq0}_{\star}=\frac{-1}{4}((\delta g^{\m\a}) g^{\n\b}(F_{\m\n}\star F_{\a\b})
+g^{\m\a}(\delta g^{\n\b})(F_{\m\n}\star F_{\a\b}))
$ and second case $\t^{oi}=0\Rightarrow$
$\delta\mathcal{L}^{\t^{0i}=0}_{\star}=
\frac{-1}{4}(F_{\m\n}\star \delta g^{\m\a}\star g^{\n\b}\star F_{\a\b}
+F_{\m\n}\star g^{\m\a}\star\delta g^{\n\b}\star F_{\a\b})$
%
%
%
%
so we get to
\bea &&
\mathbf{T}^{\t^{0i}\neq0}_{\lambda\kappa}=:\frac{\p}{\p\sqrt{-g}}S_{\star}(\sqrt{-g},g_{\lambda\kappa},
\mathcal{L}^{\t^{0i}\neq0}_{\star})
+g^{\m\n}S_\star(F_{\m\kappa},\ F_{\n\lambda})\nonumber\\&&
\mathbf{T}^{\t^{0i}=0}_{\lambda\kappa}=\frac{\p}{\p\sqrt{-g}}S_{\star}(\sqrt{-g},g_{\lambda\kappa},
\mathcal{L}^{\t^{0i}=0}_{\star})
+\frac{\p}{\p\sqrt{-g}}
S_{\star}(g^{\n\b},F_{\lambda\n},\sqrt{-g},F_{\kappa\b})
\eea
and all $\mathbf{T}_{\m\n}$ are symmetric tensors.

\textbf{Discussion}

In this work we could drive the action of quantum electrodynamics in noncommutative curved space time.
Also we find the motion equation of
electrodynamics fields.
We also assumed that $\theta^{0i}\neq0$ so the momentum conjugate of
$g^{\m\n}$ does not exhibit and metric term did not participate with star product
in Lagrangian density. additionally
we construct the typical energy-momentum tensor for two cases of noncommutativity from general way and we show that
these are symmetric tensors.

\end{document}